\documentclass{icrc2009}

\usepackage{graphicx}   % for including figures
\usepackage[font=footnotesize]{subfig} % subfig.sty for a double column floating figure using two subfigures
\usepackage{fixltx2e}
\usepackage{url}

\newcommand{\shorttitle}[1]%
{\markboth{Proceedings of the 31\MakeLowercase{$^{st}$} ICRC, {\L}\'{o}d\'{z} 2009}{#1} }
\newcommand{\etal}{\MakeLowercase{\textit{et al. }}} % "et al."

\begin{document}
\title{Performance of the MAGIC telescopes in stereoscopic mode}

\author{\IEEEauthorblockN{P. Colin\IEEEauthorrefmark{1},
			  D. Borla Tridon\IEEEauthorrefmark{1},
			  E. Carmona\IEEEauthorrefmark{1},
			  F. De Sabata\IEEEauthorrefmark{2},
              M. Gaug\IEEEauthorrefmark{3},
              S. Klepser\IEEEauthorrefmark{6},\\
			  S. Lombardi\IEEEauthorrefmark{4},
			  P. Majumdar\IEEEauthorrefmark{5},
			  A. Moralejo\IEEEauthorrefmark{6},
              V. Scalzotto\IEEEauthorrefmark{4},
			  J. Sitarek\IEEEauthorrefmark{1}\\
              for the MAGIC collaboration}
                            \\
\IEEEauthorblockA{\IEEEauthorrefmark{1} Max-Planck-Institut f\"ur Physik, Munich, Germany}
\IEEEauthorblockA{\IEEEauthorrefmark{2} Dipartimento di Fisica dell'Universit$\grave{a}$ di Udine
and INFN sez. di Trieste, Italy}
\IEEEauthorblockA{\IEEEauthorrefmark{3} Instituto de Astrofisica de Canarias, La Laguna, Spain}
\IEEEauthorblockA{\IEEEauthorrefmark{4} Dipartimento di Fisica, Universit$\grave{a}$ di Padova and
INFN sez. di Padova, Italy}
\IEEEauthorblockA{\IEEEauthorrefmark{5} Deutsches Elektronen-Synchrotron (DESY) Zeuthen, Germany}
\IEEEauthorblockA{\IEEEauthorrefmark{6} Institut de Física d'Altes Energies, Barcelona, Spain}}

% please write the preseter's name and short title (3-4 words maximum)
%    which will appear at the header of the even pages.
\shorttitle{Colin \etal Performance of the MAGIC telescopes}
\maketitle

\begin{abstract}
The MAGIC $\gamma$-ray observatory has recently been upgraded by a second 17\,m-diameter
imaging atmospheric Cherenkov telescope at a distance of 85\,m from the first one.
Simultaneous observation of air showers with the two MAGIC telescopes (stereoscopic mode)
will improve the reconstruction of the shower axis and solve the ambiguity in the impact
point occurring in single-telescope mode. Also, the stereo observation
will result in a better angular resolution, energy estimation and cosmic-ray background
rejection. It is expected that the sensitivity of MAGIC improves significantly
over the full energy range (60\,GeV - 20\,TeV). Here, we present the performance estimated
from Monte Carlo simulations.
\end{abstract}

\begin{IEEEkeywords}
 MAGIC phase-II, VHE $\gamma$-ray, Cherenkov telescope
\end{IEEEkeywords}

\section{The MAGIC Telescopes}
The MAGIC telescopes are a system of two 17\,m-diameter imaging atmospheric Cherenkov telescopes (IACT) used for very high energy (VHE) $\gamma$-ray astronomy.
The first telescope, MAGIC-I, is in operation since 2004. The second one, MAGIC-II,
was built in 2008 at a distance of 85\,m from MAGIC-I. It is currently under commissioning and scientific observation should start within a few months \cite{Cortina09}.

The structure of the two telescopes is almost identical, based on a reinforced carbon fiber frame supporting 236\,m$^2$ of mirrors. The two cameras, however, are quite different. MAGIC-I has a 3.5\,$^\circ$ field of view camera composed of two types of pixels with different sizes.
397 small pixels (0.1$^\circ$ diameter) cover a central hexagonal region and 180 larger pixels (0.2$^\circ$) cover a surrounding ring.
MAGIC-II has an improved camera \cite{Borla09} equipped uniformly with 1039 small pixels
(0.1$^\circ$) with a better photo-detection efficiency, and covering a field of view of 3.5\,$^\circ$.
The trigger region is limited to a central region of 2\,$^\circ$ diameter for
MAGIC-I and 2.5\,$^\circ$ for MAGIC-II.
Both telescopes have a 2\,GSample/s digitization data acquisition system.

The two MAGIC telescopes can be operated independently or in stereoscopic mode.
The stereoscopic observation mode allows a more precise reconstruction of the shower parameters as well as a stronger suppression of the hadronic showers and other background events.
Here we present the performance in stereoscopic mode (MAGIC phase-II) and compare
it to the performance of MAGIC in phase-I using only the MAGIC-I telescope.

\section{Monte Carlo Simulations}
The performance of the MAGIC telescopes is estimated from Monte Carlo simulations.
The simulation program of MAGIC is divided into three stages.
The CORSIKA \cite{Heck98} program simulates the air showers initiated by either
VHE $\gamma$-rays or hadrons. Here, we used the CORSIKA version 6.019,
the EGS4 code for electromagnetic shower generation
and QGSJET-II and FLUKA for high and low energy hadronic interactions respectively.
The atmospheric model is based on studies of total mass density as a function of the height. The second stage of the simulation accounts for the Cherenkov light
absorption and scattering in the atmosphere and then performs the reflection of these photons on the mirror dish. Finally, the last stage simulates the behavior of the MAGIC photomultipliers, the trigger system and the data acquisition electronics.
Pulse shapes, noise levels and gain fluctuations obtained from the MAGIC-I data have been implemented in the MAGIC-I simulation. For MAGIC-II, results from laboratory measurement are implemented.

For the study presented here, we simulated $3\times10^6$ $\gamma$-rays with an energy distribution ranging from 10\,GeV to 30\,TeV and following a power law with a
spectral index of -1.6, and impact point within 300\,m to the center of the array.
The simulated $\gamma$-ray source is point-like and located at the center of the camera.
For the background simulation, we generated $2\times10^7$ proton showers (re-used 160 times) with an energy distribution ranging from 30\,GeV to 30\,TeV with a spectral index of -1.78 over a $1200\times1200$\,m$^2$ area and a 5$^\circ$-radius view cone.
For both $\gamma$-rays and protons, the telescope pointing ranges from 5$^\circ$ to 30$^\circ$ in zenith distance and from 0$^\circ$ to 360$^\circ$ in azimuth.

The energy distributions have been chosen harder than the typical $\gamma$-ray
sources in order to increase the number of event at high energy. Nevertheless, in the analysis,
each event is weighted as a function of its energy to reproduce the spectral index of a realistic source ($\sim$$-2.5$ for the Crab nebula) and of the CR background ($\sim$$-2.7$ for protons). The analysis of simulated events is performed with standard MAGIC Analysis and Reconstruction Software (MARS) \cite{Moralejo09} which can analyze both single-telescope and stereo data.

\section{Analysis and Results}

\subsection{Image parametrization}
Cherenkov images of each telescope are calibrated, extracted and parameterized independently. The algorithms used here, are those developed for the MAGIC-I data analysis. The image cleaning method is the two-threshold method using time information \cite{Aliu09} with a high threshold of 6\,photo-electrons within a time window of $\pm$4.5\,ns, and a low threshold of 3\,photo-electrons with time constraint of 1.5\,ns.
For each image, a set of parameters (Hillas parameters \cite{Hillas85} and others) is obtained. In a IACT array, the Hillas parameters of multiple images of the same shower are generally combined in order to reduce the number of parameters. However, here
we preferred to keep the parameters separated as they are only twice and the 2 cameras
have different designs.

\subsection{Angular resolution}
In single-telescope mode, the shower direction is reconstructed with the DISP method\,\cite{Fomin94} which determines the distance between the image centroid and the shower direction as a function of the image parameters. Discrimination between the two possible directions at a distance DISP from the centroid along the major image axis is performed thanks to the image asymmetry.
With two telescopes, the primary direction and the ground impact point can be reconstructed by stereoscopy with the major axis of the 2 images (intersection of the two planes represented by the axes of the two images).

\begin{figure}[!t]
 \centering
 \includegraphics[width=6.5cm]{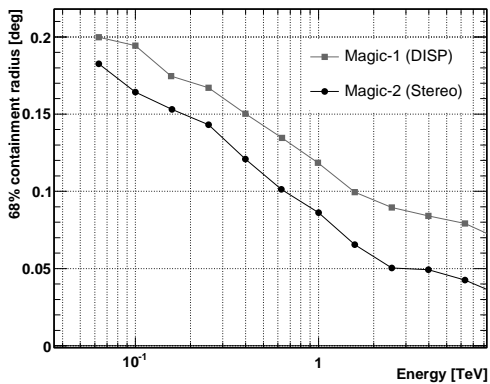}
 \caption{Angular resolution of the MAGIC observatory in single mode (MAGIC-I)
 and in stereoscopic mode.}
 \label{fig1}
\end{figure}

There are several ways to define the angular resolution of a $\gamma$-ray observatory. Here, we consider the radius of the circle centred on the simulated source, containing 68\% of the reconstructed events. Only events reconstructed at less than 0.4$^\circ$ are considered in this calculation. The others are considered as lost in the background. With the DISP method most of these events correspond to the case in which the sense of propagation has been wrongly determined ($\sim$20\%). Thus, the wrongly oriented events do not affect the angular resolution, although they result in a reduction of the effective collection area after cuts.
In stereoscopic mode, only events with nearly parallel images can be lost.

Figure~1 shows the 68\%-containment radius for $\gamma$-rays as a function of the energy for the DISP method with MAGIC-I data and for the stereo reconstruction method.
The performance of the stereo reconstruction depends on the angle between the two shower-image axes. For angles below $\sim$30$^\circ$, the DISP method is generally better than the stereo reconstruction. Selection of events with an angle between the shower-image axes above 30$^\circ$ was done for the stereo angular resolution curve shown Figure~1.
Our analysis is still under development but the final reconstructed direction would be a combination of the DISP method applied on MAGIC-I and MAGIC-II data and the stereo reconstruction. Here we simply use the stereo reconstruction for events with an angle between image axes above 30$^\circ$ and the DISP method for the others.

\subsection{Energy resolution}

\begin{figure}[!t]
 \centering
 \includegraphics[width=6.5cm]{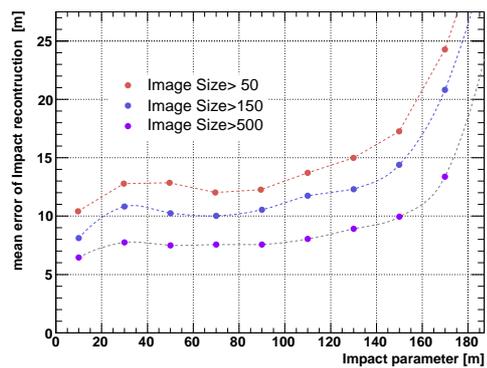}
 \caption{Mean error on the impact parameter as a function of the impact parameter for different size cuts.}
 \label{fig2}
\end{figure}

\begin{figure*}[!t]
   \centerline{
   \subfloat[Case I]{\includegraphics[width=6.5cm]{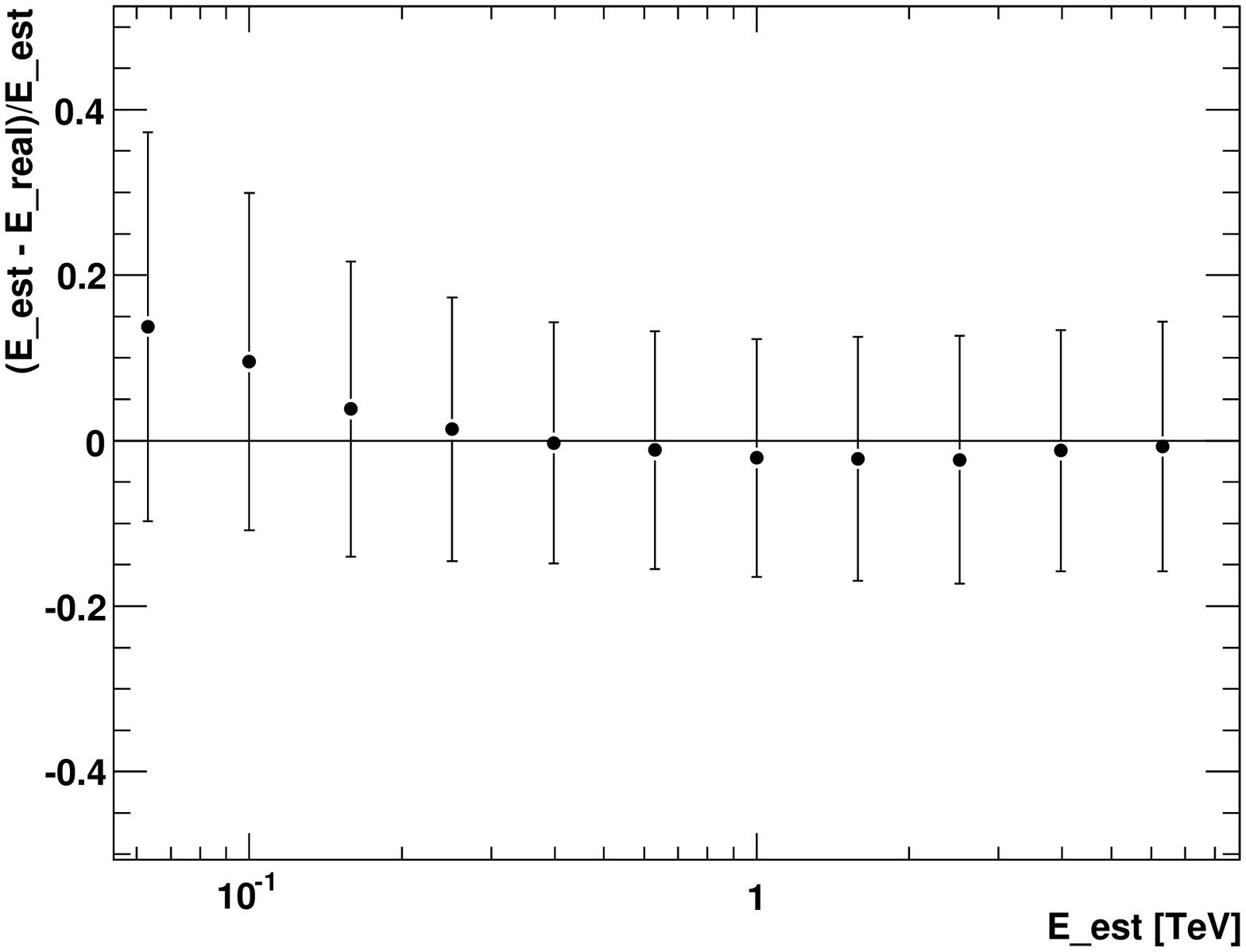}
   \label{sub_fig3a}}
   \hfil
   \subfloat[Case II]{\includegraphics[width=6.5cm]{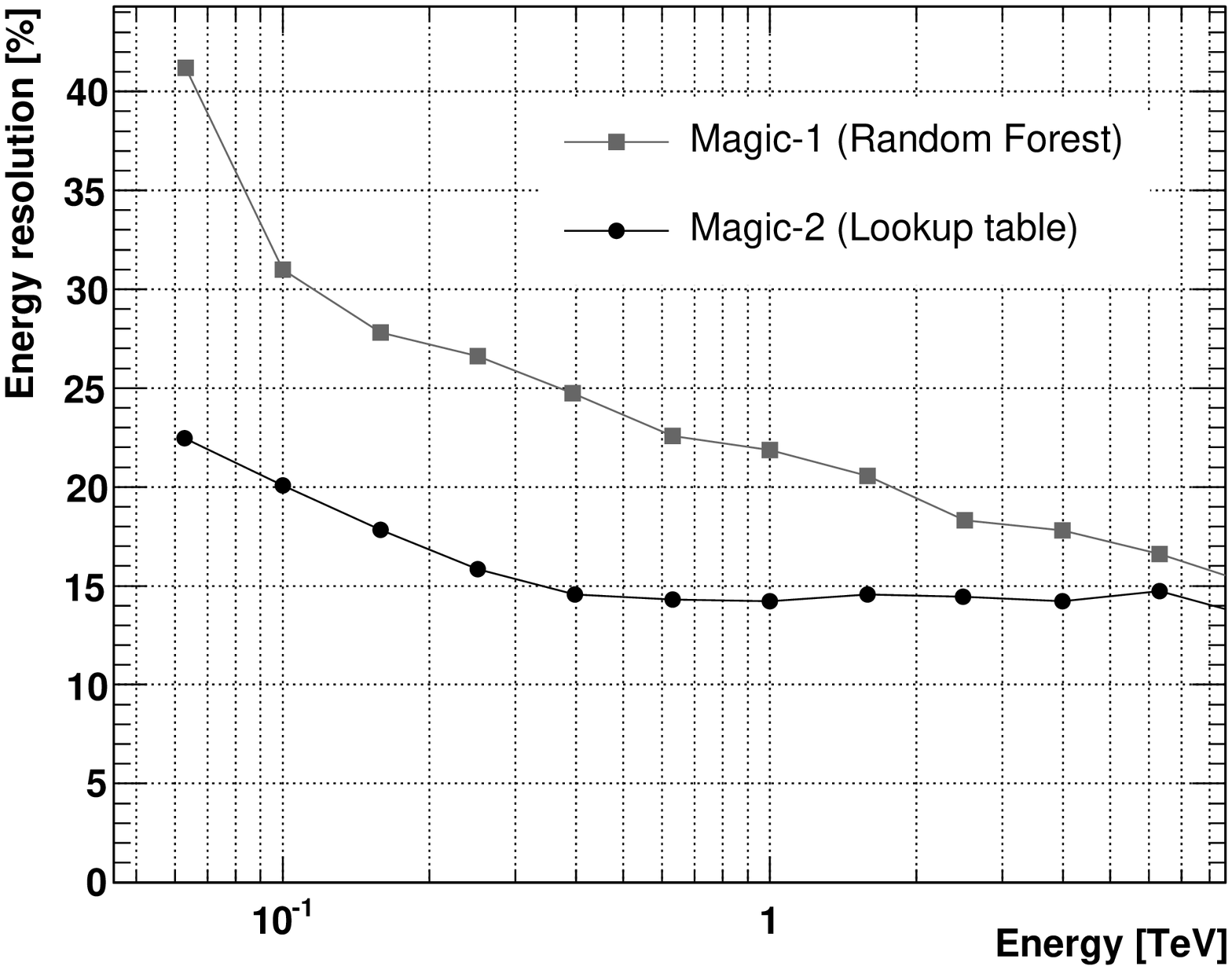} \label{sub_fig3b}}
            }
   \caption{Energy reconstruction of stereo data with lookup tables. Left-hand and right-hand panels show respectively the energy bias and the energy resolution as a function of the estimated energy. In the second plot, energy resolution obtained with stand alone MAGIC-I (RF method) is also shown for comparison.}
   \label{double_fig}
 \end{figure*}

The total number of photo-electron in an image (image size) is strongly related to
the primary energy and to the distance between the shower axis and the telescope (impact parameter). The energy reconstruction performance depends strongly on the impact
parameter reconstruction. The stereoscopic method allows to reconstruct the ground impact point (Figure~2) and also the altitude of the shower maximum ($H_{max}$) related to the distance between the shower direction and the image centroids.
In stereo-data analysis, for each telescope an energy is reconstructed using lookup tables based on the image size, impact parameter, $H_{max}$ and distance to zenith.
Combination of these two energy estimations provides the final reconstructed energy as well as a parameter describing the compatibility between these two energies.

In single-telescope mode, the impact parameter and $H_{max}$ cannot be properly reconstructed. The energy reconstruction is generally performed with the Random Forest (RF) technique \cite{Fomin94} which is an optimized event classification tool trained with a subset of $\gamma$-ray simulation.

Figure~3b compares the performance of the energy reconstruction of $\gamma$-rays with the lookup-table method applied on the stereo data and with the RF method using MAGIC-I data.
Although the lookup-table method is simpler than the RF method, it provides a better energy resolution because the 3D-parameters of the shower are well recontracted by stereoscopy. The improvement is particularly interesting at low energies.
A RF method using stereo data is under development and should improve the energy resolution even more. The results presented Figure~3b are obtained without any assumption on the source position (for both curves). It is possible to improve the energy resolution in stand-alone mode by assuming the source position (this is generally done for the spectral analysis of point-like source with MAGIC-I).
In stereoscopic mode, such assumption provides negligible improvement as the impact parameter and $H_{max}$ are already well reconstructed (see figure~2). Spectral analysis performance of extended or poorly localized sources will strongly improve with the stereo-observation.

\begin{figure}[!t]
 \centering
 \includegraphics[width=6.5cm]{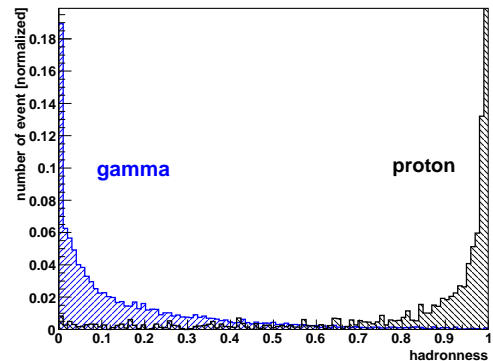}
 \caption{Hadronness distribution for $\gamma$-rays and protons with
 a reconstructed energy above 100\,GeV.}
 \label{fig4}
\end{figure}

\subsection{Gamma/hadron separation}

The $\gamma$-ray/hadron separation is performed with the RF technique.
A RF is trained to distinguish $\gamma$-ray and proton events with a set of
Monte Carlo simulations.
After training, any event can be classified by the RF, providing a parameter, called hadronness, distributed between 0 and 1. Values closer to 0 mean that
the event is more $\gamma$-ray-like and values closer to 1 mean that it is more hadron-like.
The cosmic-ray background is rejected mainly by a cut on the hadronness.
Figure~4 shows the hadronness distribution for protons and $\gamma$-rays.
As expected, the proton distribution peaks at 1 whereas the $\gamma$-ray
distribution peaks at 0. The quality factor of the selection with this cut ($Q=\epsilon_\gamma/\sqrt(\epsilon_{p})$ with $\epsilon_\gamma$ and
$\epsilon_{p}$ respectively the fraction of $\gamma$-rays and protons passing the cut)
is $Q=3.1$ at 100\,GeV and $Q=6.5$ at 500\,GeV.
The other hadrons are more easily rejected than protons, so most of the cosmic rays
passing the selection cuts are protons. The background rate can be fairly estimated with only proton simulations.

\begin{figure}[!t]
 \centering
 \includegraphics[width=6.5cm]{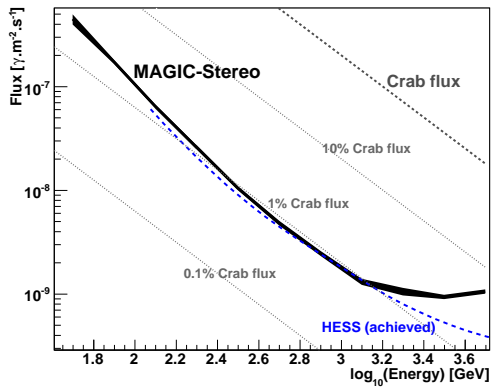}
 \caption{Sensitivity of the MAGIC telescopes in stereoscopic mode compared to
 the Crab Nebula spectrum and to the performance achieved by HESS.}
 \label{fig5}
\end{figure}

\subsection{Sensitivity}
The improvement of the $\gamma$-ray/hadron separation and energy reconstruction from the single-telescope mode to two-telescope system improves strongly the sensitivity and
spectral analysis performance of MAGIC.
The angular resolution improvement also reduces the background for point-like source
improving the sensitivity for this kind of source.
Figure~5 shows the minimal integral flux of a point-like source above a given energy for a 5-standard-deviation detection, with an excess of at least 10 events representing more than 5\% of the residual background, in 50\,h of observation with
the MAGIC telescopes in stereoscopic mode, as a function of this energy.
The current version of MAGIC-I reaches at the best a sensitivity of 1.6$\%$ of the Crab nebula flux. In stereoscopic mode, MAGIC should achieve a sensitivity better than 1$\%$.
This sensitivity is comparable to the performance of VERITAS \cite{Holder08} and HESS \cite{Becherini09} above 150\,GeV. Between 50\,GeV and 150\,GeV, MAGIC have the best sensitivity of the current IACT.
Above few TeV, the MAGIC performance suffers of the relatively small trigger region of the MAGIC-I camera.

\section{Conclusion}
In few months, the MAGIC telescopes will start observation in stereoscopic mode.
Monte Carlo simulations of MAGIC show that the angular resolution, the energy resolution, the $\gamma$-ray/hadron separation and the sensitivity will strongly improve.
In 50\,h, MAGIC should detect point-like sources with a flux lower than 1\% of the Crab nebula flux. This sensitivity is comparable to the performance of other current IACT but with an energy threshold about 2 times lower ($\sim$60\,GeV). Moreover the performances shown here are obtained with a preliminary stereo-data analysis that may probably be improved.

\section{Acknowledgements}
We thank the Instituto de Astrofisica de Canarias for
the excellent working conditions at the Observatorio del
Roque de los Muchachos in La Palma. The support of
the German BMBF and MPG, the Italian INFN, and
Spanish MCINN is gratefully acknowledged. This work
was also supported by ETH Research Grant TH 34/043,
by the Polish MNiSzW Grant N N203 390834, and by
the YIP of the Helmholtz Gemeinschaft.

%% see \section{Examples of \LaTeX\  instructions} and \subsection{Figures}
% \begin{figure}[!t]
%  \centering
%  \includegraphics[width=2.5in]{fig01}
%  \caption{Simple figure example}
%  \label{simp_fig}
% \end{figure}

%% see \section{Examples of \LaTeX\  instructions}  \subsection{Tables}
%  \begin{table}[!h]
%  \caption{A Simple Example Table}
%  \label{table_simple}
%  \centering
%  \begin{tabular}{|c|c|c|}
%  \hline
%   one  &  two & three \\
%   \hline
%    1.0 & 2.0 & 3.0\footnotemark \\
%    1.0 & 2.0 & 3.0 \\
%    1.0 & 2.0 & 3.0 \\
%    1.0 & 2.0 & 3.0 \\
%  \hline
%  \end{tabular}
%  \end{table}
%  \footnotetext{footnote inside table}

%% see \section{Examples of \LaTeX\  instructions} and \subsection{Figures}
%% An example of a double column floating figure using two subfigures.
%% The double column figure must be placed in the source text file
%%         within the text of the previous page
% \begin{figure*}[!t]
%   \centerline{\subfloat[Case I]{\includegraphics[width=2.5in]{fig02} %\label{sub_fig1}}
%              \hfil
%              \subfloat[Case II]{\includegraphics[width=2.5in]{fig03} %\label{sub_fig2}}
%            }
%   \caption{An example of a double column floating figure using two %sub figures.
%            The double column figure must be placed in the source text
%            file within the text of the previous page.}
%   \label{double_fig}
% \end{figure*}


\begin{thebibliography}{99}
   \bibitem{Cortina09}  J.~Cortina et al., 31$^{st}$ ICRC, {\L}\'{o}d\'{z} 2009 (arXiv:0907.0843).
   \bibitem{Borla09}   D.~Borla Tridon et al., 31$^{st}$ ICRC, {\L}\'{o}d\'{z} 2009 (arXiv:0907.0843).
   \bibitem{Heck98}    D. Heck et al., Technical Report FZKA, Forschungszentrum Karlsruhe, 1998.
   \bibitem{Moralejo09}  A.~Moralejo et al., 31$^{st}$ ICRC, {\L}\'{o}d\'{z} 2009 (arXiv:0907.0843).
   \bibitem{Aliu09}   E.~Aliu et al., Astropart. Phys. 30, 2009, 293
   \bibitem{Hillas85}   A. M. Hillas. In F. C. Jones, editor, 19$^{th}$ ICRC, La Jolla 1985, v3 445.
   \bibitem{Fomin94}   V. P. Fomin et al., Astropart. Phys. 2, 1994, 137
   \bibitem{Holder08}  J. Holder et al., AIP Conf. Proc., 1085, 2008, 657
   \bibitem{Becherini09} Y. Becherini, 44$^{th}$ rencontres de Moriond, La Thuile, 2009
%   \bibitem{Benbow09} W. Benbow, 44$^{\`eme}$ rencontres de Moriond, La Thuile, 2009

  \end{thebibliography}
\end{document}